\documentclass[prd,twocolumn]{revtex4}
\usepackage{amsfonts}
\usepackage{amsmath}
\usepackage{amssymb}
\usepackage{graphicx}

\providecommand{\U}[1]{\protect\rule{.1in}{.1in}}
\newcommand{\ba}{\begin{eqnarray}}
\newcommand{\ea}{\end{eqnarray}}
\def\beq{\begin{equation}}
\def\eeq{\end{equation}}

\begin{document}

\begin{titlepage}
\begin{flushright}
ACT-01/08 \\
CERN-TH-PH/2008-072\\
MIFP-08/07 \\
April 2008
\end{flushright}

\vspace*{1cm}

\begin{centering}
{\large {\bf Derivation of a Vacuum Refractive Index in a Stringy Space-Time Foam Model}}
\vspace{1cm}

{\bf John Ellis}$^a$, {\bf N. E. Mavromatos}$^b$,
{\bf D.V. Nanopoulos}$^{c,d,e}$

\end{centering}

\vspace{1cm}

\begin{centering}

{\bf Abstract}

\end{centering}

\vspace{1cm}

It has been suggested that energetic photons propagating {\it in vacuo} should experience a
non-trivial refractive index due to the foamy structure of space-time induced
by quantum-gravitational fluctuations. The sensitivity of recent astrophysical observations,
particularly of AGN Mk501 by the MAGIC Collaboration, approaches the Planck scale for
a refractive index depending linearly on the photon energy. We present here a new
derivation of this quantum-gravitational vacuum refraction index, based on a stringy
analogue of the interaction of a photon with internal degrees of freedom in a conventional
medium. We model the space-time foam as a gas of D-particles in the bulk space-time 
of a higher-dimensional cosmology where
the observable Universe is a D3-brane. The interaction of an open string representing a
photon with a D-particle stretches and excites the string, which subsequently decays and
re-emits the photon with a time delay that increases linearly with the photon energy and is related to stringy uncertainty principles.
We relate this derivation to other descriptions of the quantum-gravitational
refractive index {\it in vacuo}.

\vspace{2cm}
\begin{flushleft}
$^a$ Theory Division, Physics Department, CERN, CH-1211, Geneva 23, Switzerland. \\
$^b$ King's College
London, Department of Physics, Theoretical Physics, Strand, London WC2R 2LS, U.K. \\
$^c$ Department of Physics,
Texas A \& M University, College Station,
TX~77843-4242, U.S.A. \\
$^d$ Astroparticle Physics Group, Houston
Advanced Research Center (HARC), Mitchell Campus,
Woodlands, TX 77381, U.S.A. \\
$^e$ Academy of Athens,
Chair of Theoretical Physics,
Division of Natural Sciences, 28~Panepistimiou Avenue,
Athens 10679, Greece.
\end{flushleft}
\end{titlepage}

\section{Introduction\label{sec:1}}

Over a decade ago, we argued~\cite{aemn,aemns,mitsou} that quantum-gravitational effects 
might cause photons
of different energies to propagate at different velocities {\it in vacuo}, i.e., that quantum gravity
might induce a vacuum refractive index. Our suggestion was based on an analysis of
quantum-gravitational fluctuations in the space-time background - `space-time foam' - in a
formulation of string theory that used the Liouville field to compensate for departures from the usual
description of string vacua based on conformal field theory~\cite{horizons}. 
We also argued~\cite{aemns,mitsou} 
that the most sensitive
probes of this possibility would be provided by distant astrophysical sources producing energetic
photons in short bursts, such as gamma-ray bursters and active galactic nuclei. Subsequently, the
possibility of a vacuum refractive index has also been suggested in other theoretical frameworks (see, e.g.,~\cite{Myers}).

In parallel, increasingly sensitive and rigorous tests of the universality of the velocity of
light based on observations of photons of different energies emitted by different
astrophysical sources have been reported. The most sensitive such probe has recently been
reported by the MAGIC Collaboration, in an analysis of their data on
the 20-minute long flare of AGN Markarian (Mk) 501 observed on July 9 2006~\cite{MAGIC}, performed
together with the present authors, A.~Sakharov and E.~Sarkisyan-Grinbaum~\cite{MAGIC2}. If the vacuum
refractive index $n$ varies linearly with the photon energy: $n \sim (E/M_{\rm QG\gamma})$, 
the sensitivity of
this test approaches the Planck scale: $M_{\rm QG\gamma} \sim {\hat m}_P \equiv 1/\sqrt{8 \pi G_N}$ .

A very strong constraint on any such refractive index for electrons is provided by astrophysical observations of synchrotron radiation from Crab Nebula~\cite{crab,crab2}. These measurements 
constrain any analogous parameter for electrons, $M_{\rm QGe}$, to be several orders of magnitude larger than the Planck scale, but do not impose stringent constraints
on photon dispersion. We also note that local effective lagrangian models of quantum-gravity-induced 
modified dispersion, such as the modified quantum electrodynamics of Myers and 
Pospelov~\cite{Myers}, necessarily exhibit birefringence (i.e., different propagation speeds for the two photon helicities) as a result of the unique structure of the corresponding dimension-five local operators. However, ultraviolet (UV) radiation measurements from distant galaxies~\cite{uv} and UV/optical polarization measurements of light from  Gamma Ray Bursters~\cite{grb} rule out
birefringence unless it is induced at a scale beyond the Planck mass. Therefore, any model
of refraction in space-time foam that exhibits effects at the level of the MAGIC sensitivity
should exhibit three specific properties:
(i) photons should exhibit a modified \emph{subluminal}
dispersion relation with Lorentz-violating corrections that grow linearly with $E/M_{\rm QG\gamma}$, where $M_{\rm QG\gamma}$ is close to the Planck scale, (ii) the medium should not refract electrons, so as to avoid the synchrotron-radiation 
constraints~\cite{crab}, and (iii) the coupling of the photons to the medium must be independent of photon polarization, so as not to have birefringence. A model with all these properties has been suggested by us some years ago~\cite{emnw,ems}, and is based on a D-brane model for space-time foam.

We emphasize that a refractive index that grows with the energy of the photon is not characteristic
of the dispersion relations induced by classical gravity backgrounds. It is
well-known~\cite{hath,scharn,latorre,shore} that vacuum-polarization effects in quantum
electrodynamics in non-trivial vacua, such as finite-temperature plasmas, or between two
parallel plates, or in other restricted spaces, or in constant-curvature and cosmological
backgrounds, lead to a non-trivial refractive index for photons. For example, the loop effects of
virtual electron-positron pairs in a Lorentz-violating background such as a finite-temperature plasma
or a space with a boundary do affect the photon dispersion relation. However, these
corrections to the velocity of light are all inversely proportional to (powers of) the photon energy,
whereas quantum-gravity effects are expected to grow with energy as one approaches the
Planck energy~\cite{aemn,aemns}. Moreover, field-theoretic calculations generally yield birefringence
effects~\cite{shore}, precisely because they may be formulated in the language of low-energy
effective field theory. However, birefringence is absent in the string-inspired modified dispersion
relations arising in our non-critical string description of space-time foam~\cite{aemn,horizons,mitsou},
precisely because it {\it cannot} be described in the language of low-energy effective field theory.
Moreover, in our approach the Lorentz-violating refractive index also violates the equivalence
principle, in the sense that different particle species are expected to propagate at different
speeds. Specifically, our approach predicts that there should be {\it no} modification of the
propagation of energetic electrons or other charged particles, whereas it might also be
present for energetic particles without internal quantum numbers, such as Majorana
neutrinos~\cite{emnv}, though not necessarily with the same coefficient $M_{QG}$ as for photons.
The absence of any effect for electrons is consistent with the stringent upper limits on a
possible Lorentz-violating refractive index for electrons derived from considerations of
synchrotron radiation in the Crab Nebula~\cite{crab,crab2}.

The increasing stringency of experimental constraints on possible quantum-gravitational
refractive indices {\it in vacuo}~\cite{MAGIC2} prompts us to present here a new and conceptually simple
derivation of a refractive index for photons {\it in vacuo}, which displays explicitly why
no birefringence and no effects for electrons are expected in our intrinsically stringy
formulation of space-time foam. We derive the effect by analogy with the familiar
derivation of a refractive index for photons propagating in a material medium with non-trivial
optical properties~\cite{feynman}, which is based on the interaction of the electromagnetic
wave with the electrons in the medium, which absorb and re-emit the photons and may be treated 
(approximately) as simple-harmonic oscillators. 

Likewise, here we formulate the
quantum-gravitational medium as a set of simple-harmonic oscillators that absorb and re-emit
photons. Specifically, we model the oscillators as D0-branes (D-particles) `flashing' on and
off in a space-time background represented by a D3-brane in Type-1A string theory~\cite{emnw,ems}. 
We discuss here the microscopic dynamics of the D-particles during the photon capture and 
re-emission process, utilising the flux forces that are characteristic of D-brane 
dynamics~\cite{polchinski}. Flux conservation results in the creation of intermediate string states stretching between the D-particles and 
the D3-brane world. Their decay produces the outgoing wave corresponding to the re-emitted photon
in the familiar atomic case. In contrast to field-theoretic treatments,
there are no advanced waves, so causality is preserved and such processes cause time delays proportional to the incident photon energy. The flux forces are analogous to the restoring force for the electron in the simple-harmonic oscillator model for the refractive index~\cite{feynman}. We note
additionally that  there is a formal analogy between this recoil/capture problem and that of strings 
in a constant external electric field~\cite{seibergwitten,sussk1,sussk2}, and thereby with space-time 
non-commutativity~\cite{aemnuncert,szabo,lengthier}. The r\^ole of the electric field intensity is played here by the recoil velocity of the D-particle defect during the process~\cite{recoil,szabo}.

In the next
Section of this paper, we review the familiar derivation of the refractive index in a medium, 
and in the following Section we present our D-brane formulation of space-time 
foam~\cite{emnw,ems}. In a following
Section, we derive the non-trivial refractive index as a consequence of photon
absorption and re-emission by the D-particles using stringy methods, and then demonstrate the
consistenty of our results with the stringy space-time uncertainty principles~\cite{yoneya}.
Finally, we discuss briefly the relation of this approach to our previous derivation of this phenomenon using non-critical strings, which makes the connection between space-time 
uncertainty and non-commutativity in the target space. We leave a detailed
description for a lengthier paper~\cite{lengthier}.

\section{Review of the Derivation of a Refractive Index in a Conventional Medium}

We now review the familiar derivation of a refractive index for photon propagation through a medium in standard electrodynamics~\cite{feynman}. This illustrates the crucial similarities to, and
differences from, the string-induced case discussed in the next Section.
The underlying physics is the interaction of the photon with electrons in the medium, 
specifically absorption in the material and the excitation of internal electron degrees of
freedom that one may model (approximately) as simple-harmonic oscillators. After a characteristic
time delay related to the inverse of the oscillator frequency, the photon is re-emitted with the same
energy as it possessed initially.

We consider~\cite{feynman} the electrons, of mass $m$, as forced simple-harmonic oscillators with a resonant frequency $\omega_0$, subject to the force $F$ exerted by an
oscillating external electric field of frequency $\omega$: $F= e E_0 e^{i \omega t}$, where $e$ 
is the electron charge.  The corresponding equation of motion is:
\begin{equation}
m \left(\frac{d^2}{dt^2} x + \omega_0^2 x \right) = e E_0 e^{i \omega t}
\end{equation}
Assuming for concreteness a plate of thickness $\Delta z$, representing the medium through which an electromagnetic wave travels, where $z $ is perpendicular to $x$,
one can compute in a standard way the electric field $E_a$ produced by the excited atoms:
\begin{equation}
 E_a = -\frac{e n_e}{\epsilon_0 c}i \frac{e E_0}{m (\omega^2 - \omega_0^2)}e^{i\omega (t - z)} ,
\label{elfield}
\end{equation}
where $\epsilon_0$ is the dielectric constant of the vacuum and $n_e$ is the area density of 
electrons in the medium (plate), which is given by $n_e = \rho_e \Delta z$,
where $\rho_e$ is the volume density of electrons.
We next recall that light propagates through a medium with a refractive index $n$ with a
speed $c/n$, causing a delay $\Delta t$ in traversing the distance $\Delta z$, given by:
$\Delta t = (n - 1)\Delta z/c$. Representing the electric field before and
after passing through the plate as
$E_{\rm before} = E_0 e^{i\omega (t - z/c)}$ and 
$E_{\rm after} = E_0 e^{i\omega (t - z/c - (n-1)\Delta z/c)}$, in the case
of small deviations from the vacuum refractive index
we have: $E_{\rm after} \simeq E_0 e^{i \omega(t - z/c)} - i[\omega(n -1)\Delta z/c]
E_0 e^{i\omega(t-z/c)}$.
The last term on the right-hand-side of this relation is just the field $E_a$ produced to after
the plate by the oscillating electrons.
We then obtain from (\ref{elfield}):
\begin{equation}
(n - 1)\Delta z = \frac{n_e e^2}{2\epsilon_0 m (\omega_0^2 - \omega^2)},
\end{equation}
and hence the following formula for the refractive index in a conventional medium:
\begin{equation}
   n = 1 + \frac{{\rho_e}_e e^2 }{2\epsilon_0 m (\omega_0^2 - \omega^2)}.
   \label{refrordinary}
   \end{equation}
We see in (\ref{refrordinary}) that the refractive index in an ordinary medium is inversely proportional to (the square of) the frequency $\omega$ of light, as long as it smaller than the oscillator
frequency, where the refractive index diverges. 

Similar situations characterise the refractive index
induced by classical curvature effects in space-time, which are also inversely proportional to (some power of) the light frequency~\cite{hath,scharn,latorre,shore}.
This situation should be contrasted to the case of a (stringy) quantum-gravity medium, where
the induced gravitational refractive index is proportional to the light frequency (photon energy)
as we discuss below.

If the couplings of the two polarizations of the photon to the electrons in the medium are different, 
the phenomenon of birefringence emerges, namely different refractive indices for the two 
polarizations. Moreover, we see from (\ref{refrordinary}) that the propagation of light is
subluminal if the frequency (energy) of the photon $\omega < \omega_0$, whereas it is 
superluminal for higher frequencies (energies)~\cite{feynman}.  
This reflects the fact that the phase shift induced for the scattered light can be either positive or negative, but there such a superluminal refractive index causes no issue with causality,
since the speed at which information may be sent is still subluminal.

\section{A D-Brane Model for Space-Time Foam\label{sec:string}}

\begin{figure}[ht]
  \includegraphics[width=5cm]{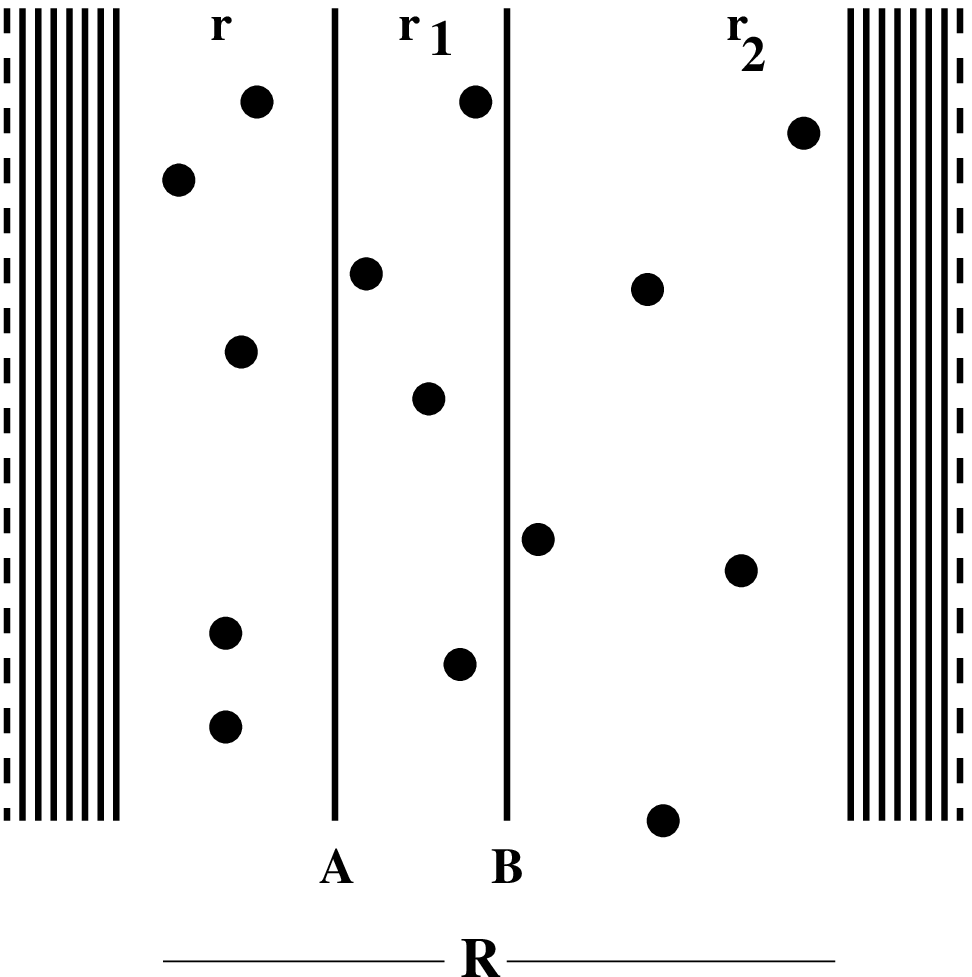} \hfill \includegraphics[width=7cm]{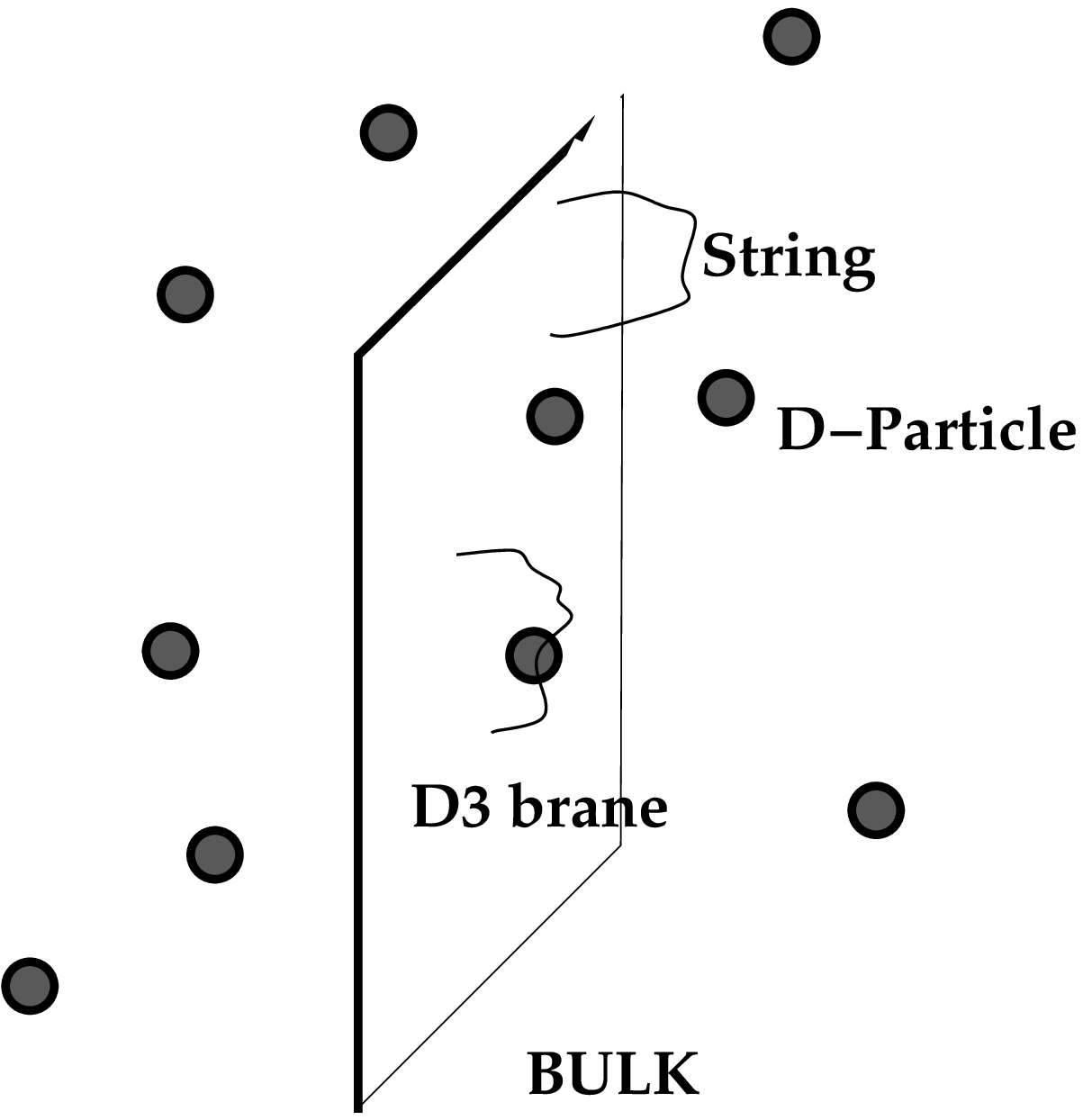}
\caption{\it A D-brane model of space-time foam in the context of Type-1A string theory.
The model consists of appropriate stacks (upper panel) of D-branes, some of which are moving
in a higher-dimensional bulk space-time, which is punctured by point-like D-brane defects (D-particles).
Thanks to relative motions between the D3-brane describing our Universe (lower panel) and these
D-particles in the bulk, the latter cross the brane world and appear to an observer on the D3-brane as space-time foam defects that `flash' on and off. The effect is `classical' from the bulk space-time
viewpoint, but appears quantum-mechanical from the viewpoint of an observer on the D3-brane.
Photons are represented as open strings on the D3-brane, and interact with these defects via
absorption and re-emission, generating a non-trivial refractive index. Charged particles do not
interact with the D-particle foam, because they cannot be absorbed by the uncharged
D-particles.}
\label{dfoam:fig}
\end{figure}

After this introductory material, we now describe our model of the quantum-gravitational
refractive index in a stringy model of space-time foam.

We constructed in~\cite{emnw} a string-inspired D-brane
model of space-time foam possessing realistic cosmological properties.
As shown in Fig.~\ref{dfoam:fig}, it consists of a ten-dimensional bulk space-time bounded
by two eight-dimensional orientifold planes. Thanks to their special reflective properties, these
orientifolds act as boundaries of the ninth dimension. The bulk space-time also contains two stacks
of eight-dimensional branes, and the entire structure is compactified to three spatial
dimensions. The bulk space is punctured by point-like D0-branes (D-particles), which are allowed in
the framework of Type-1A string theory (a T-dual of Type-1 strings~\cite{schwarz}), as
considered in~\cite{emnw} and here. The ground-state energy (dark energy or cosmological
constant) vanishes if the D-branes are stationary relative to one another.

However, conventional Big-Bang cosmology may be modelled in this framework by
postulating a collision between two of the D-branes from the original stack, causing a cosmic
catastrophe that can regarded as the initial state in a non-equilibrium cosmology~\cite{brany}.
After the collision, the D-branes recoil, and we assume that currently they
are moving slowly back towards the stack of branes from which they
emanated. As a result of this motion, the population of D-particles in the bulk cross the
D-brane worlds and interact with the stringy matter particles moving on them.
To an observer on the D-brane, the space-time defects appear to be `flashing' on and off.

Since this model involves eight-dimensional D-branes, it requires an appropriate scheme
for compactification to three spatial dimensions, e.g., by using manifolds with non-trivial
higher-dimensional magnetic fluxes (unrelated to conventional magnetic fields).
The different couplings of fermions
and bosons to such external fields break target-space supersymmetry, and the
consequent induced mass splittings~\cite{bachas,gravanis} between partner fermionic
and bosonic excitations on the D-brane world is proportional to the intensity
of the flux field. In this way, one may obtain phenomenologically realistic mass
splittings in the excitation spectrum (at the TeV or some higher energy scale), as a
result of supersymmetry obstruction. This model may lead to a value of the dark-energy
contribution to the energy budget of the observable Universe which is in agreement with current
observations. For details of this and other aspects of the model,
we refer the reader to the relevant literature~\cite{emnw,mavsark,recoil2}.

\section{A D-Brane Model for the Vacuum Refractive Index\label{sec:index}}

\begin{figure}[ht]
  \includegraphics[width=8cm]{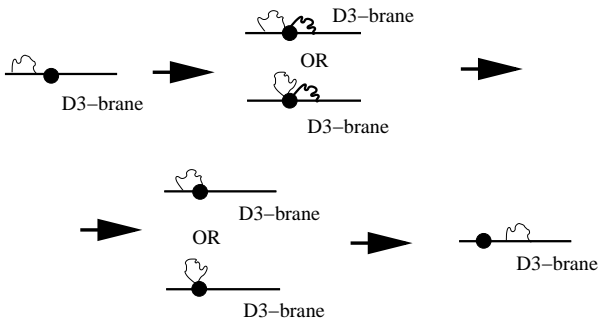}
\caption{\it Schematic view of the capture by a D-particle of an open-string state
propagating on a D3-brane world, representing a photon. The possible intermediate string states
(indicated by the thick wavy lines) that are created by the capture of the end(s) of the open
photon string by the D-particle, stretch between the D-particle and the brane world.
They oscillate in size between $0$ and $\sim \alpha ' p^0$, where $p^0$ is the energy of the
incident photon. They subsequently decay by emitting outgoing photon waves during the
re-emission process. The intermediate string state provides the restoring force that keeps
the D-particle in its ground-state configuration after the scattering of the photon string state.}
\label{fig:restoring}
\end{figure}

We now consider the non-trivial interaction of an open string representing a photon with a
D0-brane traversing our D3-brane world. This interaction is described schematically in
Fig.~\ref{fig:restoring}. It involves the capture of
the open string by the D-particle defect, which becomes excited, and
subsequently re-emits the photon.
This process is very analogous with the mechanism for generating a refractive index in a
material medium via the interaction of a photon with an electron in an atom.
Since there are no charged D-particle excitations, the
conservation of electric charge prevents charged excitations, such as electrons, from participating
in such processes. For this reason, in the model of~\cite{emnw} only photons~\cite{ems} and possibly neutrinos~\cite{mavsark} may interact non-trivially with the D-brane foam.
It is this non-universality of the D-particle foam that allows, as already mentioned, the avoidance of the stringent synchrotron radiation constraints of~\cite{crab}, which would otherwise exclude time delays proportional to  the photon energy that are suppressed by a single power of a mass scale of the
order of the reduced Planck mass ${\hat m}_P$. This is close to the sensitivity exhibited by the
MAGIC data on the AGN Mk501~\cite{MAGIC,MAGIC2}, and
leaves open the possibility of a refractive index that depends linearly on the photon energy and
is suppressed by a single power of the string scale, that might be accessible to observation.

We now describe in more detail the analogous stringy physics underlying our model of
the the refractive index generated by space-time foam. We note first that it is
adequate to consider the D-particles in the foam as static, compared to the photon.
This is because the ends of the open string representing the photon move on the D3-brane world
with (essentially) the speed of light in a conventional vacuum: $c \to 1$ in our units. In
contrast, the characteristic velocities of D-particles relative to the brane world
are necessarily lower and
as discussed in \cite{brany}, in order to reproduce the spectrum of primordial density fluctuations
in this brane-world model, the speed of the D3-brane representing our Universe should be smaller
than $10^{-4}c$.

Because of the conservation of the characteristic fluxes of D-branes~\cite{polchinski}
due to stringy symmetries, there are no isolated D-particles. They must always be 
connected by strings stretched to either a D3-brane or another D-particle~\cite{szabo}.
The tension of these flux-carrying strings corresponds in this stringy quantum-gravity
model to the atomic interactions in conventional media.
It is because of these interactions that the analogy holds of the D-particles with the electrons in the 
simple-harmonic-oscillator model~\cite{feynman} for the conventional refractive index. The 
flux-carrying interactions play the r\^ole of the restoring force in that model.

When the end(s) of the open-string photon state is (are) attached to the D-particle,
as in Fig.~\ref{fig:restoring}, an intermediate
string state is formed, thanks to the above-mentioned flux conservation.
This stores the incident energy $p^0$ of the photon as potential energy, and
is stretched between the D-particle and the D3-brane. The string
grows in size to a length $L$ that is determined by the requirement of energy minimization, as
we now discuss.
We assume that, in addition to simply stretching to a length $L$, the intermediate string state may also
acquire $N$ internal oscillator excitations. This implies that the energy may be written in the form:
\begin{equation}
   p^0 = \frac{L}{\alpha '} + \frac{N}{L}.
\label{noscill}
\end{equation}
Minimizing the right-hand side determines $N$, which then is substituted back to the equation to yield the required maximal length $L$:
\begin{equation}\label{lmax}
L_{\rm max} = \frac{1}{2}\alpha ' p^0.
\end{equation}
Since the end of the stretched string state that remains attached to the D3-brane moves with the speed
of light in (normal) vacuo, $c=1$, the time taken for the intermediate string state first to grow to this
maximal length and then to shrink again to its minimal size is:
\begin{equation}
  \Delta t \sim  \alpha ' p^0
\label{delayonecapture}
\end{equation}
This describes the time delay experienced by a photon propagating through D-particle foam,
whereby the formation of the intermediate composite string state between D-particles and the photon
shown in Fig.~\ref{fig:restoring} resembles the excitation of internal degrees of freedom in a
conventional medium~\cite{feynman}. We now note several significant features of this result. 

$\bullet$ Photon propagation in our D-particle model of quantum-gravitational space-time foam is
necessarily {\it subluminal}, avoiding any potential problems with gravitational {\v C}erenkov
radiation. 

$\bullet$ The time delay (\ref{delayonecapture}) is independent of the
photon polarization, and hence the capture process of fig.~\ref{fig:restoring} leads to 
{\it no birefringence}.
For this reason, our D-particle foam model avoids the stringent constraints coming from astrophysical observations~\cite{uv,grb}. 

$\bullet$ The derivation of the delay 
(\ref{delayonecapture}) does not rely on a local effective Lagrangian description of the effect. 
This is an important feature of our stringy approach, differentiating it from models that attribute time delays to modified dispersion relations obtained from a local effective lagrangian, such as the 
modified QED Model of~\cite{Myers}.

$\bullet$ The effect is absent for particles carrying conserved charges, such as electrons, because
there are no charged D-particles in our model. Thus, the speeds of energetic particles do not
become universal in the high-energy limit, causing a breakdown of the equivalence principle, as
well as Lorentz invariance. An energetic graviton would propagate subluminally, like a
photon, but the quantum-gravitational refractive index might differ in magnitude.
In a supersymmetric extension of this model, the photino (a
Majorana particle) would experience an effect similar to that on the photon. If neutrinos are
Majorana particles, they might also propagate subluminally at high energies, but not
necessarily at the same speed as photons with the same energies.

%The (quantum) oscillations of the intermediate string state will produce a series of outgoing wave-packets, with attenuating amplitudes, which will correspond to the re-emission process of the photon after capture. The presence of the stretched string state, which carries the characteristic flux of the D-brane interactions, provides the restoring force, necessary to keep the D-particle in its position after scattering with the photon. The situation may be thought of as the stringy/brany analogue of the restoring force in situations in local field theories of photons propagating in media with non trivial refractive indices, as discussed by Feynman.

The above discussion was in the Dirichlet picture, describing the attachment of the ends of the strings on D-branes.  In the Neumann picture, the above situation is described by the scattering of wave-packets of string states~\cite{sussk2}, where again an intermediate stretched string state is formed, when the packets lie close to each other,. This grows in size from zero to a maximal length $\alpha ' p^0$, determined by the above-described  energy minimization procedure (\ref{noscill}). In such a case both ends of the intermediate string state move with the speed of light, which again gives a delay of the form (\ref{delayonecapture}) in order of magnitude.
Such delays for wave-packets are perfectly consistent with the string uncertainty principles,
as we discuss below.

We now remark that, in the case of Neumann strings in the presence of a constant electric field, with intensity $\vec E$, as was considered in~\cite{sussk2}, the time delay is modified to become
\begin{equation}
   \Delta t \sim  \frac{\alpha ' p^0}{1 -|\tilde{\vec{E}}|^2},
\label{cefdelay}
\end{equation}
where $\tilde{E}_i \equiv {E_i}/{E_{\rm c}}$ and $E_{\rm c} \equiv {1}/{2\pi \alpha '}$ is the Born-Infeld critical electric field, that characterizes open strings in an external constant electric field 
background~\cite{polchinski,seibergwitten,sussk1}. We note that the appearance of the critical 
electric field $E_{\rm c}$ is associated with the vanishing of the denominator
of the right-hand-side of (\ref{cefdelay}). This reflects the destabilization of the vacuum when the electric field intensity approaches the critical value, which was noted in \cite{burgess}.

As discussed in~\cite{szabo}, to be developed in a forthcoming work~\cite{lengthier}, 
there is an analogy between our D-particle space-time foam model and the case of a constant 
uniform electric field. The recoil of the D-particle defect during the capture process shown in 
Fig.~\ref{fig:restoring} induces a distortion
of the surrounding space-time, via a non-diagonal metric element proportional to the
the average recoil velocity $\overline u_i$ of the defect~\cite{recoil}. Formally, this plays the 
same r\^ole as a constant electric field, that is:
\begin{equation}
   \Delta t \sim  \frac{\alpha ' p^0}{1 -|\vec u|^2}~: \qquad \overline u_i = g_s\Delta k_i/M_s,
\label{cefdelayrecoil}
\end{equation}
where $\Delta k$ is the average momentum transfer during the propagation of the photon in the D-particle foam medium, $M_s=1/\sqrt{\alpha '}$ is the string scale, and $g_s$ is the string coupling.

It can be shown formally~\cite{lengthier} that the resemblance of (\ref{cefdelay}) and 
(\ref{cefdelayrecoil}) is not
a mere coincidence. Exploiting the transformation of the appropriate
boundary conditions under T-duality, it can be shown that a Dp-brane wrapped around a compact
dimension, i.e., a circle of radius $R$, and carrying a constant electric field $E$ that
points in this direction, is  T-equivalent to a D(p-1)-brane moving with
constant velocity $u = 2 \pi \alpha^\prime E$ around the dual compact dimension, i.e., one with
radius $R^\prime = \alpha^\prime /R$.

We note that, in the case of relativistic recoil velocities $|\overline u_i | \to 1$ that are attained
when the photon energy $E \sim M_s$, the time delay $\Delta t \to \infty$, which is another 
manifestation of the above-mentioned destabilization of the vacuum in the case of superluminal velocities or, equivalently, when the external `electric' field reaches its critical value.
We also note the similarity of this feature to the divergence in the conventional refractive
index (\ref{refrordinary}) when the photon frequency reaches the oscillator frequency.

The time delay (\ref{cefdelayrecoil}) is due to a single encounter of a photon with a D-particle. In case of 
space-time foam with a linear density of defects $n^*/\sqrt{\alpha '}$, i.e., $n^*$ defects per string 
length, the overall delay encountered in the propagation of the photon from the source to observation
over a distance $D$ is (assuming that the total photon energy $p^0$ is conserved on the average, i.e.,
ignoring small fluctuations due to the D-particle recoil fluctuations discussed below):
\begin{equation}
\Delta t_{\rm total} = \alpha ' p^0 n^* \frac{D}{\sqrt{\alpha '}} = \frac{p^0}{M_s} n^* D.
\label{totaldelay}
\end{equation}
The total delay (\ref{totaldelay}) may be thought of as implying~\cite{feynman} an effective subluminal refractive index $n(\omega)$ for light with frequency $\omega$ propagating in space-time.

This delay reflects the slowing down of light by medium effects. We use
$e^{i\omega t - i \vec k \cdot \vec x}$ as a plane-wave basis over which we expand a photon wave packet. A delay $\Delta t$, implies that the photon traverses a distance $\Delta L$ in time $t + \Delta t $, that is its velocity has effectively been diminished to:  $c(\omega) \equiv \frac{c}{n(\omega)} = \frac{\Delta L}{t + \Delta t} = \frac{\Delta L}{t (1  + \Delta t/t)}$, with $c=\Delta L/t$ the constant speed of light in normal vacuo. From this we obtain the subluminal refractive index:
\begin{equation}
n(\omega) - 1 = \frac{\Delta t}{t}  = c\frac{\Delta t}{\Delta L}  \propto \alpha ' p^0
\end{equation}
on account of (\ref{totaldelay}).

\section{Connections with Other Approaches}

We now comment briefly on the relationship of this derivation to other approaches to string theory.

$\bullet$ First, we observe that the above time delays are directly connected to the stringy space-time uncertainty relation~\cite{yoneya}:
\begin{equation}
\Delta x \Delta t \ge \alpha '~.
\label{stunc}
\end{equation}
To see this, we simply combine (\ref{stunc}) with the Heisenberg phase-space uncertainty
relation: $\Delta x \Delta p \ge 1$ in natural units. Identifying the momentum $p$ with the
energy $p^0$, as is appropriate for a massless particle, we see that $\Delta x \ge 1/p^0$.
Inserting this last inequality into (\ref{stunc}), we find that $\Delta t \ge \alpha^\prime p^0$,
which is consistent with our result (\ref{delayonecapture}). This is hardly surprising, since both
(\ref{stunc}) and (\ref{delayonecapture}) are essentially stringy effects, the latter being
associated with the capture process shown in Fig. \ref{fig:restoring}.

%These delays are \emph{causal}, and do not characterize local field theories in non commutative space times (the analogue of having strings in constant electric fields), which suffer from \emph{non causal effects}, due to the existence of advanced outgoing waves after the scattering.
%The fact that string theory produces only retarded waves after the scattering, and not advanced ones,
%has been explained by a detailed calculation in \cite{sussk2} in the case of the scattering of two open-string tachyons (for simplicity). There are scattering phases which are such that \emph{in string theory only delays occur}, and there are \emph{no} advanced, \emph{acausal}, signals, in contrast to local non-commutative field theories. This issue is related with the \emph{maintenance of causality} by strings, which notably is violated in the case of field theories in non-commutative space times, where a similar scattering of wave packets would result to advanced outgoing wave-packets, violating causality.

$\bullet$ The result (\ref{stunc}) has been derived previously within the context of
non-critical Liouville string theory. In that approach, the conventional critical string
framework based on conformal field theory on the world sheet is extended by
introducing a Liouville field $\phi$ that compensates for non-conformality (non-criticality).
The field $\phi$ is introduced as a dynamical renormalization scale that acquires a
time-like metric if the effective string theory goes super-critical as happens, e.g., during an
inflationary epoch in the early Universe~\cite{brany}, or because the D3-brane worlds move
in the cosmological framework used here~\cite{emnw}, 
or in the D-particle recoil process~\cite{recoil,recoil2} discussed above.
We have further argued that the (appropriately normalized
logarithm of the zero mode of the) time-like Liouville field can be
identified with the time variable $t$~\cite{emn}. 
In the case of the D-particle recoil process discussed above,
the recoil velocity of the D-particle causes the departure from criticality, inducing a non-diagonal
term in the effective background metric `felt' by the energetic photon that is proportional to its
energy $p^0$. This induces, in turn, a time delay of the form (\ref{delayonecapture}).
The new features here are that we introduce a new microscopic description of the capture
and recoil process that is formulated within the language of critical string theory and is
closely related to the familiar description of refraction in a medium.
A more complete description of the relationship between the two approaches will be
given elsewhere~\cite{lengthier}.

$\bullet$ Within the Liouville approach to recoil~\cite{recoil}, the identification~\cite{emn} 
of the target time $t$ with the Liouville mode causes $t$ to become
an operator, after summing up world-sheet genera~\cite{emninfl}. This time
variable fails to commute with the spatial coordinate operators~\cite{aemnuncert,szabo}, 
a form of non-commutativity analogous to that for strings in external electric 
fields~\cite{seibergwitten,sussk1,sussk2}. This explains naturally the space-time uncertainty 
relation~\cite{yoneya}, which, as was discussed above, is consistent with the time delay
(\ref{delayonecapture}) that we find, which is proportional to the photon energy $p^0$. 
In our case, the r\^ole of the external electric background field is played by the recoil velocity 
$u_i$ of the space-time D-particle defect.  This relation underlies the formal similarity
between (\ref{cefdelay}) and (\ref{cefdelayrecoil}).

We shall present
a more detailed discussion of these issues in a future publication~\cite{lengthier}.

\section{Conclusions}

Just as the consideration of photons interacting with electrons provide a derivation of
the Heisenberg uncertainty relation $\Delta x \Delta p \ge 1$, we have seen that
the D-particles play the r\^ole of the electrons in in our D-brane model of quantum-gravitational 
space-time foam. In our case, the D-particle foam is used as a way of `explaining' the stringy 
uncertainty relation and space-time non-commutativity, as a result of D-particle recoil.
Thanks to this connection, MAGIC~\cite{MAGIC} and other $\gamma$-ray telescopes become
`microscopes' for probing~\cite{MAGIC2} string uncertainties and space-time non-commutativity 
in the framework of D-particle space-time foam.

\section*{Acknowledgements}

The work of J.E. and N.E.M. is partially supported by the European Union
through the Marie Curie Research and Training Network \emph{UniverseNet}
(MRTN-2006-035863) and that of D.V.N. by DOE grant DE-FG02-95ER40917.

\end{document}